\documentclass[final,1p,times]{elsarticle} 

\usepackage{graphicx}
\usepackage{amssymb} 
\usepackage{amsmath} 
\usepackage{amsthm} 
\usepackage{lineno}

\journal{Nuclear Physics A} 

\begin{document}

\begin{frontmatter} 

\title{Hydro overview}

\author[1]{Jean-Yves Ollitrault}
\author[auth2]{Fernando G. Gardim}
\address[1]{CNRS, URA2306, IPhT, Institut de physique th\'eorique de Saclay, F-91191 Gif-sur-Yvette, France}
\address[auth2]{Instituto de F\'\i sica, Universidade de S\~ao Paulo, C.P. 66318, 05315-970, S\~ao Paulo-SP, Brazil}

\begin{abstract} 
We review recent progress in applying relativistic hydrodynamics to
the modeling of heavy-ion collisions at RHIC and LHC, with emphasis on
anisotropic flow and flow fluctuations. 
\end{abstract} 

\end{frontmatter} 


\section{Introduction}

Relativistic hydrodynamics is the only first-principles
approach~\cite{Baier:2007ix} to the out-of-equilibrium dynamics of the
strongly-coupled quark-gluon plasma formed in heavy-ion collisions. 
It is a macroscopic description in which the plasma is modeled as a
continuous lump of fluid expanding into the vacuum. 
It plays an important role in understanding the soft sector of 
nucleus-nucleus collisions at RHIC and LHC.  

The goal of relativistic hydrodynamics is to describe the bulk of
particle production. A hydrodynamic calculation uses as input a model
for initial conditions --- typically, for the initial density profile
in a collision. 
The only input needed for hydrodynamics is the initial value of the
energy-momentum tensor: all other microscopic details of the initial
state are washed out by local thermalization, which is implicitly
assumed by the hydrodynamic description. 
One then solves for the fluid expansion using
equations of ideal or viscous relativistic hydrodynamics. 
Eventually, the fluid is transformed into {\it independent\/} 
hadrons\footnote{Depending on the implementation, further hadronic decays
and/or rescatterings may occur after hadronization.} 

The fluid is continuous: single-particle
distributions from the fluid can be computed (as a function of
transverse momentum $p_t$, pseudorapidity $\eta$, azimuthal angle $\phi$) 
with unlimited accuracy for a given initial condition, unlike in an
actual experiment. As we shall see later, this peculiar feature of the fluid
description is an important one. 
Many groups implement hadronization using a Monte-Carlo generator,
thus mimicking the experimental situation  where at most a few thousand
particles are observed per event\footnote{This is the most natural way
  of implementing a ``hadronic afterburner'', i.e., further 
  rescatterings after hadronization.}. This amounts to picking randomly one
element out of a thermal ensemble. The fluid is the thermal ensemble. 

In 2010, it was shown that observed azimuthal correlations between
particles separated by a gap in pseudorapidity (usually referred to as
``long-range'' correlations) are compatible with this hydrodynamic
picture~\cite{Alver:2010gr,Luzum:2010sp}, thus providing a unique
signature of collective behavior.  
In these proceedings, we review the status of this ``flow hypothesis''
in light of recent measurements. 
We then list open issues, and recent works since the last Quark Matter
conference.  

\begin{figure}
\begin{center}
 \includegraphics[width=.4\linewidth]{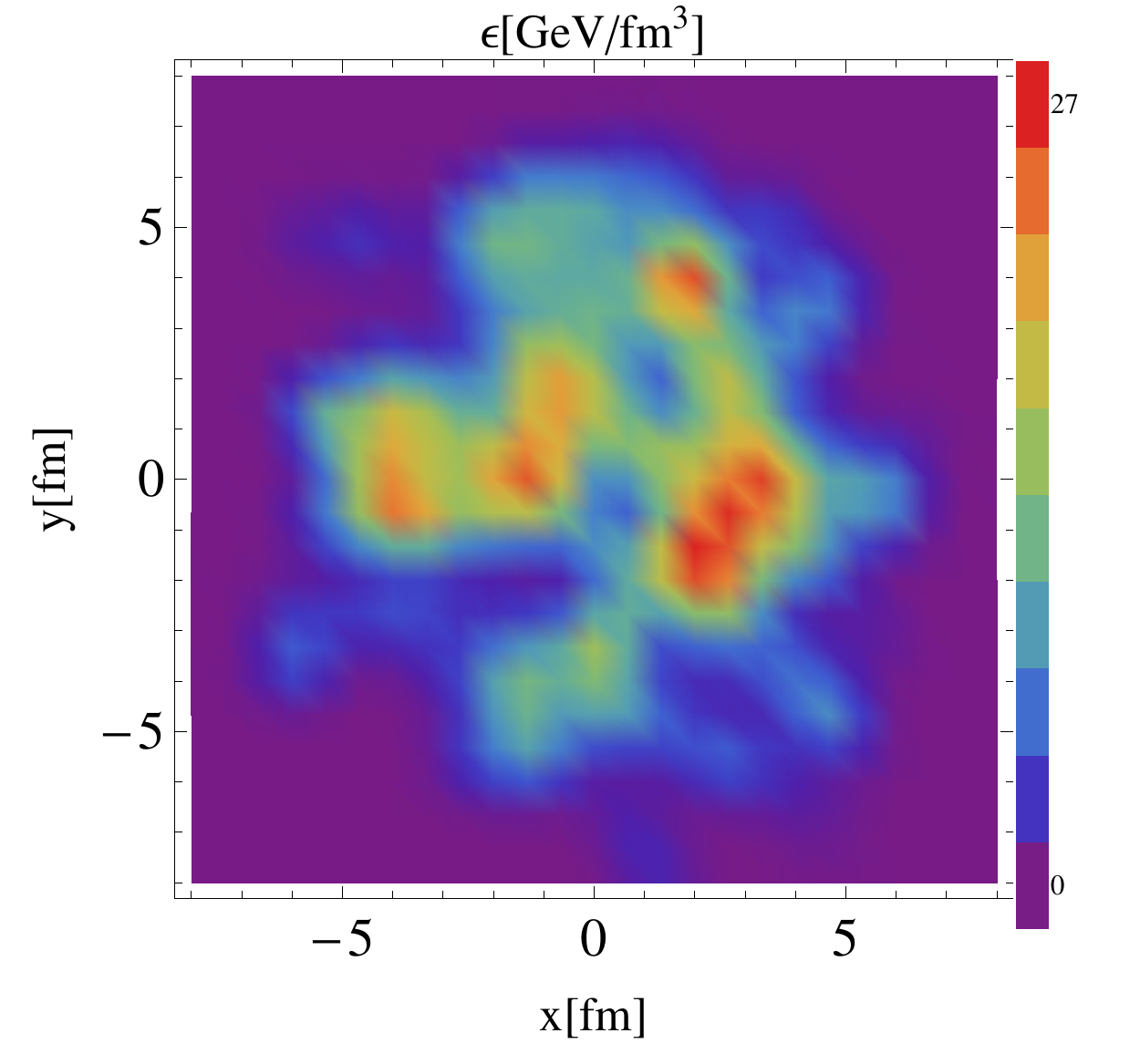}
 \includegraphics[width=.59\linewidth]{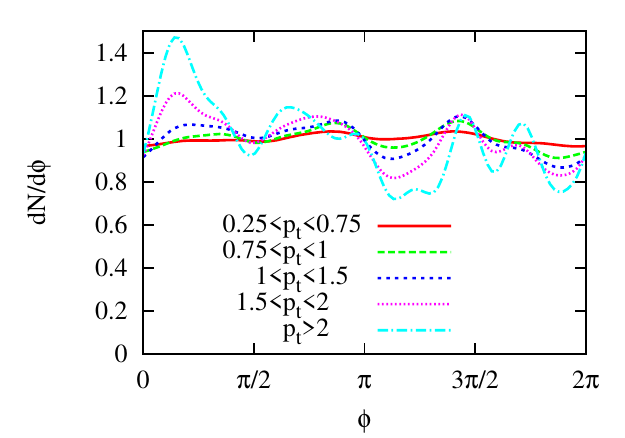}
\end{center}
\caption{(Color online) Left: initial energy density in the transverse
  plane at $z=0$   in a random central ($b=0$) Pb-Pb collision at
  2.76~TeV simulated by   Nexus. 
Right: azimuthal distribution of outgoing particles after evolving
this initial distribution through ideal hydrodynamics, in several bins
of transverse momentum $p_t$ (in units of GeV/$c$).}
\label{fig:phidist}
\end{figure}

\section{A close look at a hydro event} 
Quantum fluctuations in the wavefunctions of colliding nuclei result 
in an initial profile which is not smooth and fluctuates from event to
event~\cite{Alver:2006wh}. 
In order to understand the physics associated with these initial
fluctuations, it is instructive to look at a particular event in some
detail. We simulate a central ($b=0$) Au-Au collision using the event
generator NeXus~\cite{Drescher:2000ha}, which gives the initial 
density profile shown in Fig.~\ref{fig:phidist} (left).  
These initial conditions are then evolved through ideal
hydrodynamics~\cite{Hama:2004rr}. 
As explained above, thermalization thus transforms a single event into
a thermal ensemble. The term ``hydro event'' usually refers to this
thermal ensemble. We transform the fluid into hadrons using a
Monte-Carlo generator. This hadronization is 
repeated a few thousand times so that we can reliably calculate
ensemble-averaged quantities for the single hydro event displayed 
in Fig.~\ref{fig:phidist} (left).

A central collision between spherical nuclei is azimuthally symmetric,
except for quantum fluctuations. 
One clearly sees in Fig.~\ref{fig:phidist} (left) that fluctuations
break rotational symmetry in the transverse plane. 
Fig.~\ref{fig:phidist} (right) displays the azimuthal distribution of
charged particles in the pseudorapidity interval $|\eta|<1$ for
various bins in $p_t$. 
Anisotropies are at the $\%$ level at low $p_t$ and become stronger
and stronger as $p_t$ increases. 

The azimuthal distribution can be expressed as a Fourier series:
\begin{equation}
\label{complexfourier}
 \frac {2\pi} {N} \frac {dN}{d\phi} = \sum_{n=-\infty}^{\infty} V_n(p_t,\eta) e^{-i n \phi},
\end{equation}
where $V_n = \{e^{in\phi}\}$ is the $n$th (complex) Fourier 
coefficient, and curly brackets indicate an average over the
probability density in a single event.  
Writing $V_n=v_ne^{in\Psi_n}$, where $v_n$ is the (real) anisotropic
flow coefficient and $\Psi_n$ the corresponding phase, and using
$V_{-n}=V_n^*$ (where the superscript $^*$ denotes the complex
conjugate), this can be  rewritten as
\begin{equation}
\label{realfourier}
 \frac {2\pi} {N} \frac {dN}{d\phi} = 1 + 2\sum_{n=1}^\infty
 v_n(p_t,\eta) \cos n\left(\phi - \Psi_n(p_t,\eta)\right).
\end{equation}
Note that, for this form to describe an arbitrary distribution, both
$v_n$ and $\Psi_n$ may depend on transverse momentum $p_t$ and
pseudorapidity $\eta$.
Most models of initial conditions predict fluctuations in the form of
longitudinally-extended flux tubes~\cite{Dumitru:2008wn}. Therefore,
one expects $\Psi_n$ and $v_n$ to depend little on rapidity. 
Although this rapidity dependence is worth 
investigating~\cite{Bozek:2010vz,Florchinger:2011qf,Pang:2012he}, we
focus here on the stronger $p_t$ dependence. 

\begin{figure}
\begin{center}
 \includegraphics[width=.333\linewidth]{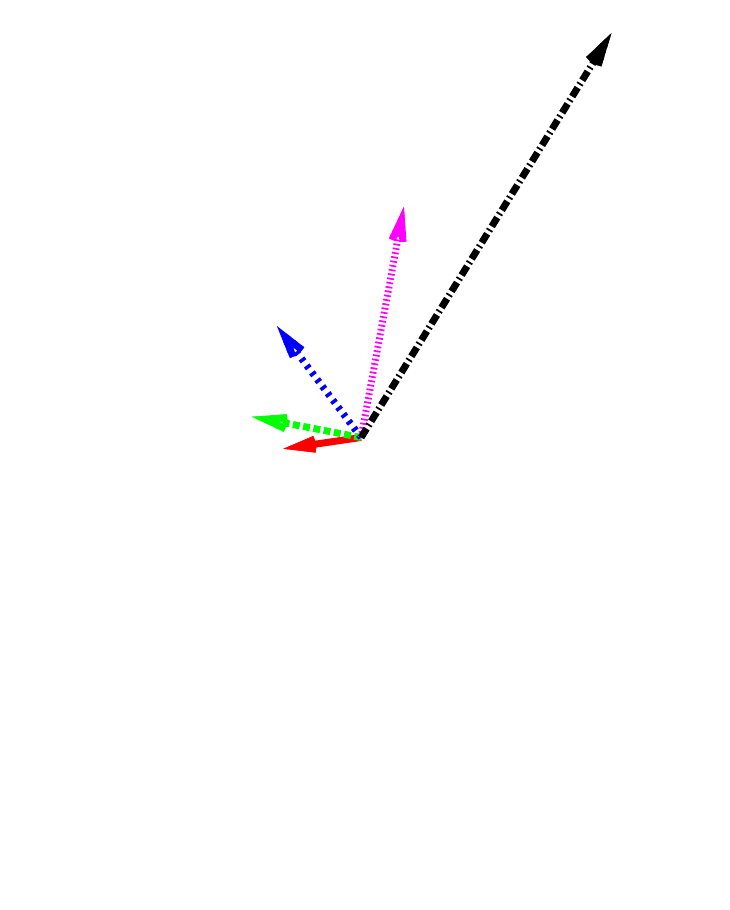}\includegraphics[width=.333\linewidth]{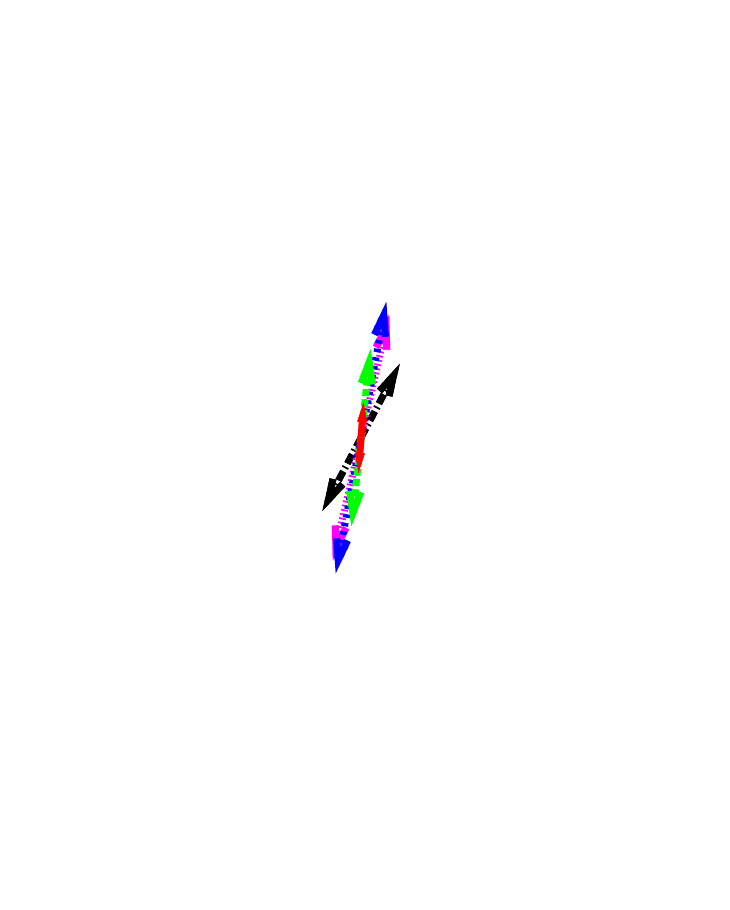}\includegraphics[width=.333\linewidth]{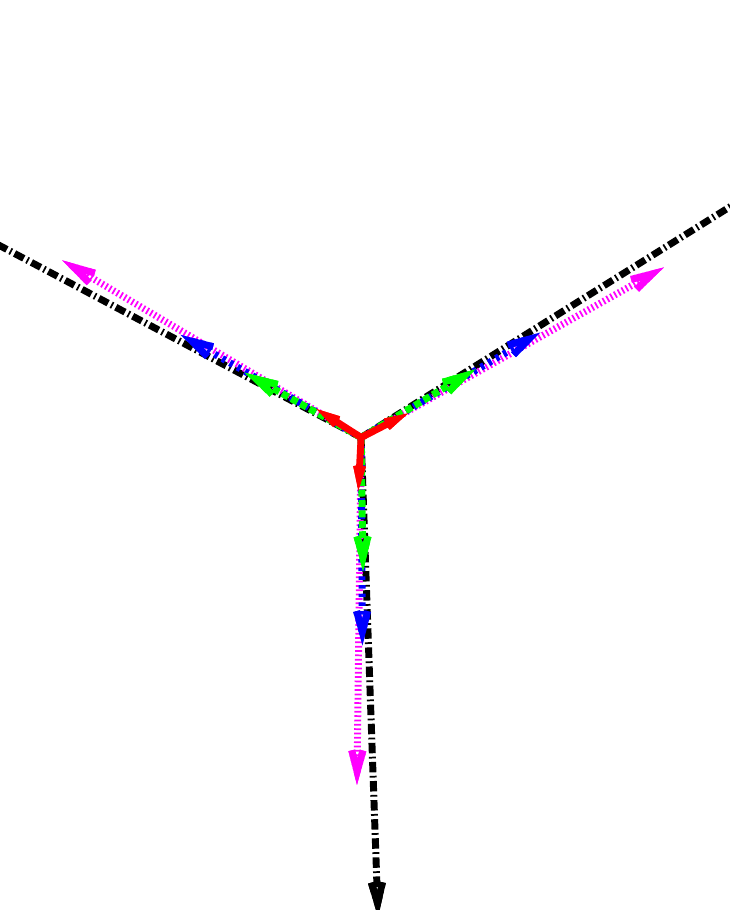}
\end{center}
\caption{(Color online) Arrows representing the magnitude and
  direction of the first 
  Fourier coefficients of the curves in Fig.~\ref{fig:phidist} (right). The
  length of the arrow is $v_n$ and its orientation is $\Psi_n$. 
  From left to right: $n=1$ (directed flow), $n=2$ (elliptic flow),
  $n=3$ (triangular flow).}
\label{fig:arrows}
\end{figure}
Note that event-by-event anisotropic flow is thus well defined in
hydrodynamics {\it only\/}, because thermalization transforms a single 
event into a thermal ensemble. 
In an actual experiment, the relative statistical error is typically
50\% for  $v_2$ (in a mid-central Pb-Pb collision analyzed by CMS or
ATLAS), and 100\% for $v_3$. 
Therefore the event-by-event $v_n$, as defined above from the
single-particle distribution, cannot be measured experimentally. 

Fig.~\ref{fig:arrows} displays the magnitude ($v_n$) and direction
($\Psi_n$) of the first three harmonics obtained by a Fourier
decomposition of the distributions in Fig.~\ref{fig:phidist}. Note
that $\Psi_n$ is defined modulo $2\pi/n$,  
and is therefore represented using $n$ arrows pointing into the
corresponding directions. 
As noted before, azimuthal anisotropies generally increase with $p_t$,
hence $v_n$ increases with $p_t$. The directions of elliptic flow
($\Psi_2$) and triangular flow ($\Psi_3$) depend slighly on
$p_t$\footnote{The particular event that we randomly picked for this
  illustration turns out to have a rather small $v_2$ and large $v_3$.}.
The $p_t$ dependence is much stronger for $\Psi_1$. This reflects the fact
that the net transverse momentum of the central rapidity slice is
expected to be close to 0, hence the integral of
$p_tV_1(p_t)=p_tv_1(p_t)e^{i\Psi_1(p_t)}$  over all particles should
be 0. One therefore typically expect that $\Psi_1$ rotates by $\pi$ as
transverse momentum goes from $0$ to 
$\infty$\footnote{The hydrodynamic
  code (NeXSPheRIO) that we use has a  small residual net transverse
  momentum due    to the choice of initial conditions, as noted in a
  previous study of   directed  flow~\cite{Gardim:2011qn}. This effect
  has not been   corrected here.}.

\section{Flow in data: a close look at 2-particle correlations} 

As explained above, event-by-event anisotropic flow is well defined in
hydrodynamics only. 
The number of particles per event in an actual experiment is too small 
to measure $V_n(p_t,\eta)$, or even the integrated $v_n$. 
Anisotropic flow can only be measured through event-averaged azimuthal
correlations between particles. 
The simplest azimuthal correlation is the pair correlation. 
One forms all possible pairs of particles and then performs a harmonic
decomposition of the distribution of the relative angle: 
\begin{equation}
V_{n\Delta} (p_t^a,
p_t^b)\equiv\langle\cos(n(\phi_a-\phi_b))\rangle=\langle
e^{in(\phi_a-\phi_b)}\rangle.
\label{expe}
\end{equation}
This quantity can be measured as a function of the transverse momenta
of both particles, labeled as $a$ and $b$, thus yielding a correlation
matrix. This correlation matrix has recently been measured by LHC 
experiments~\cite{Aamodt:2011by,Chatrchyan:2012wg,ATLAS:2012at}. 
It is symmetric by construction\footnote{ATLAS~\cite{ATLAS:2012at} uses
  a different binning in $p_t$ for the two particles in the pair, thus breaking
  the symmetry.}.

We now evaluate this quantity in hydrodynamics and show
that it is can be expressed simply in terms of anisotropic flow. 
In a single hydro event, particles are emitted independently.
Therefore 
\begin{equation}
 \label{complex} 
\left\{ e^{in(\phi^a - \phi^b)} \right\} =
\left\{e^{in\phi^a}\right\} \left\{ e^{-in\phi^b} \right\}
 = V_n^{a}V_n^{b*}=v_n^a v_n^b e^{in (\Psi_n^a - \Psi_n^b)}, 
\end{equation}
where $\{\cdots\}$ denotes an average over a single hydro event, 
The first equality in Eq.~(\ref{complex}) expresses mathematically
that particles are independent: this implies that the two-particle
correlation, {\it when written in complex form\/}, factorizes in a single
hydro event.  

The experimental quantity, Eq.~(\ref{expe}), is then obtained by
averaging over a larger number of hydro events:
\begin{equation}
\label{complexav}
V_{n\Delta} (p_t^a, p_t^b) =\left\langle V_n^{a}V_n^{b*}\right\rangle=
\left\langle v_n^a v_n^b e^{in (\Psi_n^a -
    \Psi_n^b)}\right\rangle 
\end{equation} 
Due to parity symmetry, only the real part remains after this average,
hence the cosine in Eq.~(\ref{expe}). 

From this relation alone, one can make the following general
statements about the event-averaged correlation matrix when flow is dominant: the diagonal elements must be positive, and the off-diagonal elements must satisfy a Cauchy-Schwarz inequality~\cite{Gardim:2012},
\begin{eqnarray}
V_{n\Delta}(p_t^a,p_t^a) &\geq& 0 ,\cr
\label{CS}
|V_{n\Delta}(p_t^a,p_t^b)|^2 &\leq& V_{n\Delta}(p_t^a,p_t^a)V_{n\Delta}(p_t^b,p_t^b) .
\end{eqnarray}
It is often stated that flow implies factorization. 
As shown above, factorization holds for a single hydro event. 
It implies that the second inequality in (\ref{CS}) is 
saturated, i.e., equality is achieved. 
After averaging over events, one does not in general expect
factorization to hold, therefore the 2nd inequality is generally a strict
one for $p_t^a\neq p_t^b$.  
Any violation of (\ref{CS}) is an unambiguous 
indication of the presence of non-flow correlations.

\begin{figure}
\begin{center}
 \includegraphics[width=.357\linewidth]{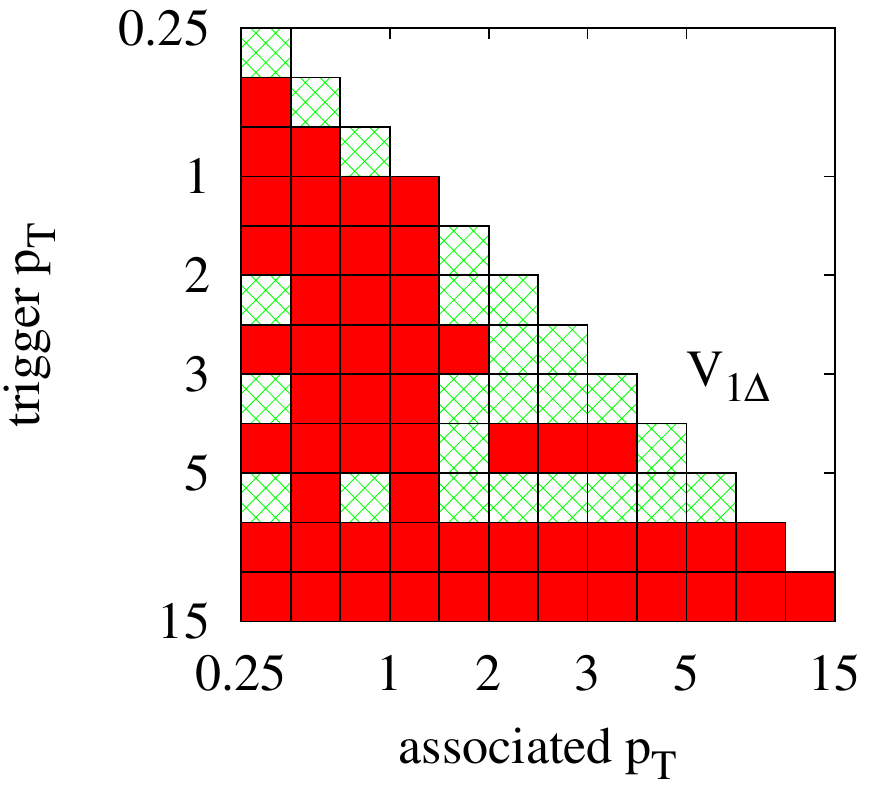}\includegraphics[width=.2856\linewidth]{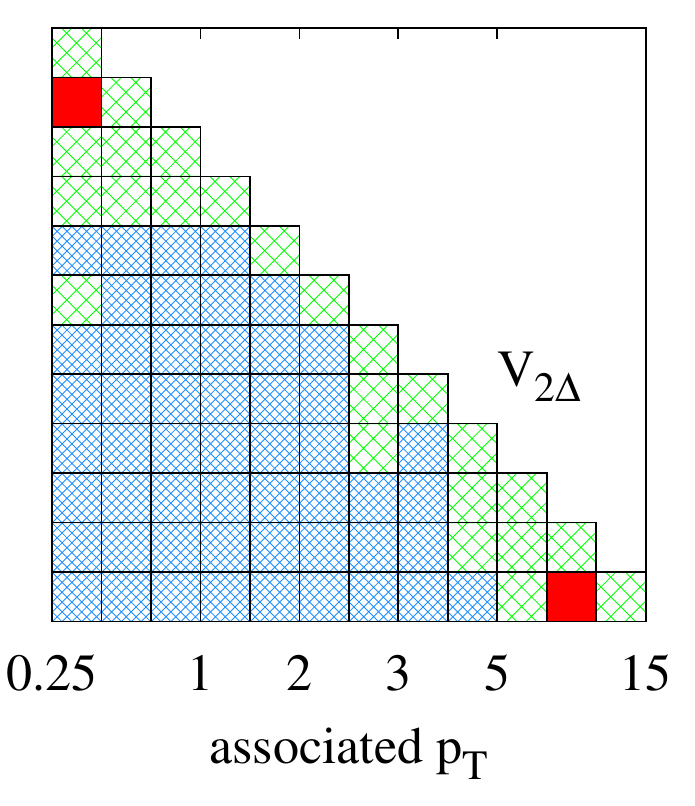}\includegraphics[width=.2856\linewidth]{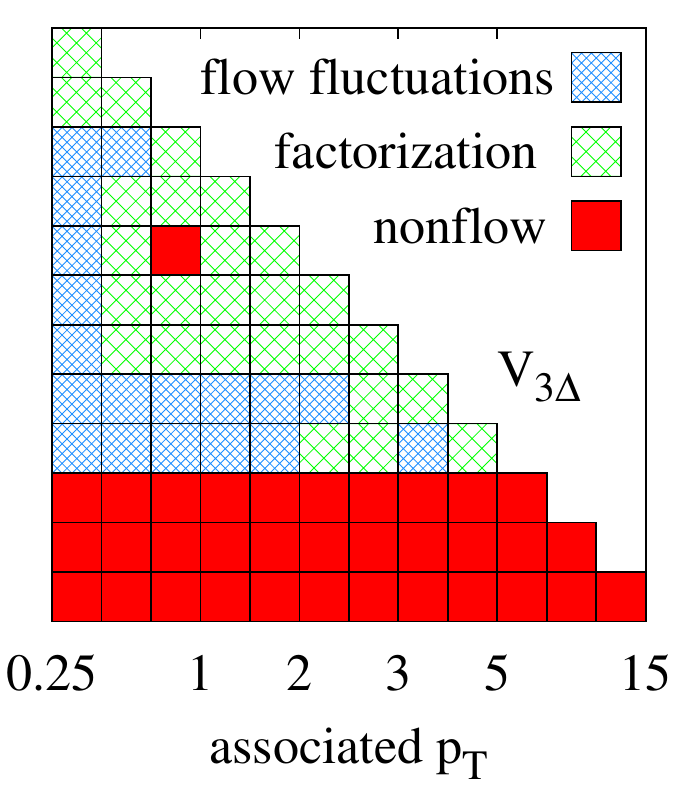}
\end{center}
\caption{(Color online) Test of inequalities (\ref{CS}) for the
  correlation matrix  $V_{n\Delta} (p_t^a,p_t^b)$ measured by 
  ALICE~\cite{Aamodt:2011by} for the 10\%  most central Pb-Pb
  collisions at 2.76~TeV. 
Green: the second inequality is an equality
  within errors. Blue: strict inequality (flow fluctuations). Red: 
  inequalities violated (correlations cannot be explained by hydrodynamics). 
From left to right: $n=1,2,3$. Only statistical errors were taken into
account. 
The boundaries of the $p_t$ bins are $0.25$,
  $0.5$, $0.75$, $1$, $1.5$, $2$, $2.5$, $3$, $4$, $5$, $6$, $8$, $15$ (in
  GeV/$c$). 
We use the standard terminology and denote by 
trigger (associated) the particle with the 
higher (lower) $p_t$, although this distinction is irrelevant here. 
}
\label{fig:matrix}
\end{figure}
Inequalities (\ref{CS}) can be directly tested on experimental data. 
We use ALICE data~\cite{Aamodt:2011by} for 0-10\% central Pb-Pb
collisions. For $n=2$ (Fig.~\ref{fig:matrix}, middle), all diagonal
elements are positive, and the Cauchy-Schwarz is verified except for two
matrix elements, where the violation is compatible with a statistical
fluctuation (the deviation is barely above 1~$\sigma$). 
The ALICE collaboration concluded from their analysis that the
correlation matrix factorizes approximately for trigger particles
below 4~GeV/$c$. However, deviations from factorization are 
clearly seen at much lower transverse momentum. It is interesting to
note that they are everywhere compatible with the general inequalities  
predicted by hydrodynamics. If hydrodynamics is a valid
approach, such deviations are a direct evidence of event-by-event
fluctuations.  

For $n=3$ (Fig.~\ref{fig:matrix}, right), diagonal elements are negative above 5~GeV, thus violating
the first inequality (\ref{CS}) and providing evidence for breakdown
of the independent-particle hypothesis underlying the hydrodynamic
picture. The observed effect can be qualitatively explained by
correlations from jets: the trigger and associated  particles are
separated by a pseudorapidity gap and cannot belong to the same
jet. But they can belong to the same pair of back-to-back jets, 
in which case their azimuthal angles are typically separated by
$\Delta\phi\simeq\pi$, thus giving a negative contribution to diagonal
elements for odd $n$. 
Below 5~GeV, all matrix elements are compatible with inequalities
(\ref{CS}), except for one point (which is again compatible with a
statistical fluctuation). As observed for $n=2$, factorization is
broken even at low $p_t$.   

Finally, for $n=1$ (Fig.~\ref{fig:matrix}, left), the inequalities (\ref{CS}) are massively
violated, both at low and high $p_t$, thus indicating that nonflow
effects are important. One of these nonflow effects is the 
correlation induced by global momentum 
conservation, which only
contributes to the first Fourier harmonic. 
After correcting for this effect, data are 
dominated by collective flow~\cite{ATLAS:2012at,Retinskaya:2012ky}. 

Comparisons between hydro and data so far only address the ``single
particle $v_n(p_t)$'', which is inferred from the correlation between
a single particle with momentum $p_t$ and all the particles in a
reference detector. Such measurements amount to averaging the
correlation matrix, Fig.~\ref{fig:matrix}, over a line or a column. 
As we have shown above, the detailed structure of the matrix contains
much more information; in particular, it provides direct insight into 
event-by-event fluctuations. 
The correlation matrix defined by Eq.~(\ref{complexav}) can be 
directly evaluated in event-by-event hydrodynamics. 
Future event-by-event hydrodynamics should address the full structure
of two-particle correlations, which can be directly compared with
experimental data. 

For the sake of brevity, we have only discussed the simplest
correlations, i.e., pair correlations. 
Much additional information is contained in higher-order 
correlations. Higher-order cumulants give direct information on the
magnitude of event-by-event flow
fluctuations~\cite{DerradideSouza:2011rp}. 
A much wider range of possibilities is opened up by mixed correlations
between event planes, which have recently been measured at the  
LHC~\cite{Jia:2012sa,Bilandzic:2012an} and can be directly evaluated 
in event-by-event hydrodynamic calculations~\cite{Qiu:2012uy}.
Such higher-order correlations are likely to play an important part 
in the near future of hydrodynamic calculations. 

\section{Recent progress and open issues}

It has long been recognized~\cite{Luzum:2008cw} that the dominant
source of uncertainty in comparisons between hydro and experimental
data lies in the modeling of initial conditions. 
The modeling of event-by-event fluctuations introduces further  
uncertainty and has triggered much recent 
activity~\cite{Alvioli:2011sk,Dumitru:2012yr,Schenke:2012hg}. 
From the point of view of hydrodynamics, it is important to understand
the hydrodynamic response to a given initial condition: for instance,
to what extent elliptic flow represents the response to the initial
eccentricity of the interaction region. This hydrodynamic response has
been studied quantitatively~\cite{Gardim:2011xv} and the importance of
nonlinear terms has been pointed out~\cite{Teaney:2012ke}. 

As for the hydrodynamic expansion itself, the dominant source of
uncertainty is the value of transport
coefficients~\cite{Song:2012ua}. Bulk viscosity was  
recently shown to have a small effect on integrated flow 
observables~\cite{Dusling:2011fd,Schaefer:2012fc}. Shear viscosity, on
the other hand, is known to reduce anisotropic flow: the
higher the harmonic, the stronger the effect~\cite{Alver:2010dn}. 
But higher harmonics also have a larger uncertainty from initial
conditions: presently, RHIC data are equally well reproduced 
with a minimal shear viscosity~\cite{Schenke:2011bn} 
or with ideal hydrodynamics~\cite{Gardim:2012yp}, depending on how
initial conditions are modeled. 
Studying ultra-central collisions at the LHC may help reducing the 
uncertainty on the shear viscosity~\cite{Luzum:2012wu}. 

The last stage of the hydrodynamic evolution is the hadronic phase. It
is not clear at present whether hadronic interactions are strong
enough for hydrodynamics to be a valid description of this phase. 
There seems to be a consensus that some hadronic interactions 
are needed in order to match spectra of identified hadrons, but that
ideal hydrodynamics fails. 
A transport calculation (hadronic afterburner) reproduces data quite 
well~\cite{Song:2011qa}. Alternatively, one can use hydrodynamics with
a non-vanishing bulk viscosity~\cite{Bozek:2012qs}. 

\begin{table}[ht]
\centering
\begin{tabular}{|l|c|c|c|c|c|}
\hline
\em Author & \em Ref.& \em initial fluctuations & \em 3+1d & \em viscous& \em afterburner\\
\hline
Dusling&\cite{Dusling:2011fd}& & &\checkmark&\\
Schenke&\cite{Schenke:2011bn}&\checkmark &\checkmark &\checkmark&\\
Derradi de Souza&\cite{DerradideSouza:2011rp}&\checkmark &\checkmark&&\\
Yan YL&\cite{Yan:2011tn}&&\checkmark &&\checkmark\\
Chaudhuri&\cite{Chaudhuri:2011pa}&\checkmark &&\checkmark&\\
Petersen&\cite{Petersen:2012qc}&\checkmark &\checkmark &&\checkmark\\
Vredevoogd&\cite{Vredevoogd:2012ui}&&\checkmark &\checkmark&\\
Shen C&\cite{Shen:2012vn}& & &\checkmark&\\
Gardim&\cite{Gardim:2012yp}&\checkmark &\checkmark &&\\
Retinskaya&\cite{Retinskaya:2012ky}& & &\checkmark &\\
Ryblewski&\cite{Ryblewski:2012rr}&&\checkmark &&\\
Bozek&\cite{Bozek:2012en}&\checkmark &\checkmark &\checkmark&\\
Nonaka&\cite{Nonaka:2012qw}&\checkmark &\checkmark &\checkmark&\\
Karpenko&\cite{Karpenko:2012yf}&&\checkmark &&\checkmark\\
Hirano&\cite{Hirano:2012kj}&\checkmark &\checkmark &&\checkmark\\
Pang LG&\cite{Pang:2012he}&\checkmark &\checkmark &&\\
Teaney&\cite{Teaney:2012ke}& & &\checkmark&\\
Song H&\cite{Song:2012tv} & & &\checkmark&\checkmark\\
Holopainen&\cite{Holopainen:2012id}&\checkmark &&&\\
Soltz&\cite{Soltz:2012rk}&&&\checkmark &\checkmark\\
Qiu Z&\cite{Qiu:2012uy}&\checkmark & &\checkmark &\\
Luzum&\cite{Luzum:2012wu}& & &\checkmark &\\
Ryu&\cite{Ryu:2012at}&\checkmark &\checkmark &\checkmark &\checkmark\\
\hline
\end{tabular}
\caption{Recent works (since the Quark Matter 2011 conference, listed
  in chronological order) containing
  numerical hydrodynamic calculations applied to heavy-ion
  collisions. We indicate by check marks whether or not calculations
  involve event-by-event fluctuations of the initial state; whether
  they are three dimensional or two-dimensional (with Bjorken
  longitudinal expansion); whether they are ideal or viscous; and
  finally, whether or not collisions are implemented in the hadronic phase
  (hadronic afterburner). 
}
\label{tbl}
\end{table}
Table~\ref{tbl} is a list of papers 
published since the last Quark Matter conference and containing
numerical hydrodynamic calculations applied to ultrarelativistic
nucleus-nucleus collisions. 
The list is by no means exhaustive. In particular, it does not include
theoretical developments: 
Progress has been made in understanding the relationship between
transport theory and hydrodynamics~\cite{Denicol:2012cn}, which is
relevant to our understanding of the hadronic phase. 
Another important development is the relativistic theory of
hydrodynamical fluctuations~\cite{Kapusta:2011gt}.
The correlations resulting from such intrinsic 
fluctuations~\cite{Bozek:2012en,Gavin:2012if}, which are typically
thought of as  ``nonflow'' correlations, could thus eventually be
studied within the framework of hydrodynamics~\cite{Springer:2012iz}.

\section{Perspectives}

Hydrodynamics has been the state-of-the-art approach to
the soft sector of nucleus-nucleus collisions for at least a decade. 
It was first used to explain the large value of elliptic flow at RHIC. 
Other azimuthal correlations, such as the ``soft
ridge''~\cite{Alver:2008aa}, were largely thought to be ``nonflow''
effects, and it took a few years to recognize~\cite{Alver:2010gr} that 
they were also naturally described by hydrodynamics. 
In these proceedings, we have pointed out for the first time that the
detailed structure  of azimuthal correlations~\cite{Aamodt:2011by} 
might be explained by hydrodynamics alone below 5~GeV. 
Hydrodynamics encompasses a wider and wider range of phenomena. 

The field of nucleus-nucleus collisions is characterized by a strong
interplay between theory and experiment. 
Most experimental measurements of anisotropic flow are biased by the 
hydrodynamic picture where flow is a single-particle observable. 
Elliptic flow is often presented as a ``single-particle'' observable, even
though all measurements are inferred from azimuthal correlations. 
Triangular flow was actually discovered~\cite{Alver:2010gr} through a
thorough analysis of these correlations. 
We advocate that experiments measure quantities which are simple and
unambiguous. The pair correlation (\ref{expe}) is a good example of
such a simple ---yet non-trivial--- measurement.
In turn, theory should directly address what is measured, i.e.,
correlations. 
This will allow for a more fruitful interplay between theory and experiment.

Hydrodynamics applied to ultrarelativistic heavy-ion collisions is an
active and lively field of research, and the recent 
measurements open up exciting new perspectives. 
The structure of pair correlations will be investigated 
as a function of transverse momentum and rapidity. 
The recently-measured correlations between event 
planes~\cite{Jia:2012sa,Bilandzic:2012an} 
will also further constrain models and tighten our knowledge of the
initial state.

\section*{Acknowledgments}
We thank Fr\'ed\'erique Grassi and Matthew Luzum for discussions. 
JYO is supported by the European Research Council under the
Advanced Investigator Grant ERC-AD-267258.

\bibliographystyle{h-elsevier3}
\bibliography{spires}
\end{document}